\documentclass[fleqn,usenatbib]{mnras}

\usepackage{newtxtext,newtxmath}

\usepackage[T1]{fontenc}
\usepackage{ae,aecompl}
\usepackage{color}

\usepackage{graphicx}	
\usepackage{amsmath}	
\usepackage{amssymb}	

\newcommand{\comment}[1]{}
\newcommand{\as}{$^{\prime\prime}$ } 
\newcommand{\red}[1]{\textcolor{red}{#1}}

\makeatletter
\AtBeginDocument{\let\red\@firstofone}
\makeatother

\title[Stereo-SCIDAR at ESO Paranal]{Optical turbulence profiling with Stereo-SCIDAR for VLT and ELT}

\author[J. Osborn et al.]{J. Osborn,$^{1}$\thanks{E-mail: james.osborn@durham.ac.uk (JO)}
R. W. Wilson$^{1}$,
M. Sarazin$^{2}$,
T. Butterley$^{1}$,
A. Chac\'{o}n$^{2}$,
F. Derie$^{2}$,
\newauthor 
O. J. D. Farley$^{1}$,
X. Haubois$^{2}$
D. Laidlaw$^{1}$,
M. LeLouarn$^{2}$,
E. Masciadri$^{3}$,
J. Milli$^{2}$,
\newauthor
J. Navarrete$^{2}$
and M. J. Townson$^{1}$,
\\
$^{1}$Centre for Advanced Instrumentation, University of Durham, Durham, UK\\
$^{2}$European Southern Observatory, Karl-Schwarzshild-Str.2, 85748 Garching bei Muenchen, Germany\\
$^{3}$INAF Osservatorio Astrofisico di Arcetri, Largo Enrico Fermi 5, I-50125 Florence, Italy
}

\date{Accepted XXX. Received YYY; in original form ZZZ}

\pubyear{2015}

\begin{document}
\label{firstpage}
\pagerange{\pageref{firstpage}--\pageref{lastpage}}
\maketitle

\begin{abstract}
Knowledge of the Earth's atmospheric optical turbulence is critical for astronomical instrumentation. \red{Not only does it enable performance verification and optimisation of existing systems but it is required for the design of future instruments.} As a minimum this includes integrated astro-atmospheric parameters such as seeing, coherence time and isoplanatic angle, but for more sophisticated systems such as wide field adaptive optics enabled instrumentation the vertical structure of the turbulence is also required.

Stereo-SCIDAR is a technique specifically designed to characterise the Earth's atmospheric turbulence with high altitude resolution and high sensitivity. Together with ESO, Durham University has commissioned a Stereo-SCIDAR instrument at Cerro Paranal, Chile, the site of the Very Large Telescope (VLT), and only 20~km from the site of the future Extremely Large Telescope (ELT). 

Here we provide results from the first 18 months of operation at ESO Paranal including instrument comparisons and atmospheric statistics. Based on a sample of 83 nights spread over 22 months covering all seasons, we find the median seeing to be 0.64\as with 50\% of the turbulence confined to an altitude below 2~km and 40\% below 600~m. The median coherence time and isoplanatic angle are found as 4.18~ms and 1.75\as respectively.

A substantial campaign of inter-instrument comparison was also undertaken to assure the validity of the data. The Stereo-SCIDAR profiles (optical turbulence strength and velocity as a function of altitude) have been compared with the Surface-Layer SLODAR, MASS-DIMM and the ECMWF weather forecast model. \red{The correlation coefficients are between 0.61 (isoplanatic angle) and 0.84 (seeing).}
\end{abstract}




\section{Introduction}

\red{Turbulence within the Earth's atmosphere imposes a limitation upon astronomical observations. Wavefront distortions blur the image and can be compensated with Adaptive Optics (AO) systems.} The future of this correction technique requires comprehensive knowledge of the dynamics of the Earth's atmosphere. This is critical for future sophisticated AO systems on existing very large and the future extremely large telescopes (ELTs). These telescopes will be sensitive to variations in turbulence altitude of the order of 100 to 500~m (see for example \citealp{Neichel2008, Basden2010, Vidal2010, Gendron2014}).

The next generation of 40~m class ELTs that are currently under construction will enable new discoveries in all areas of astronomy and push forwards the boundaries of human knowledge. \red{They will look further back in space and time to explore the early universe and shed light on currently unanswered questions such as the physical basis of dark energy and dark matter, as well as their evolution in the time scales from early universe to present time.} They will discover and characterise extra-solar planets and potentially find distant habitable worlds. \red{More details of ELT science cases can be found in \cite{ESO2009, skidmore2015}}. To fulfil these ambitious objectives these giant telescopes will be equipped with highly sophisticated AO in order to counteract the detrimental effects of the Earth's atmosphere.

SCIDAR (Scintillation Detection and Ranging) \citep{Vernin73} is a technique often used for profiling the Earth's atmospheric turbulence. Stereo-SCIDAR is an extension of the SCIDAR technique. Stereo-SCIDAR is a sensitive, high-altitude resolution turbulence monitor capable of returning the vertical profile of Earth's optical turbulence strength and velocity in real-time. The Stereo-SCIDAR instrument has been described previously \citep{Shepherd13}. It has been routinely and reliably used at the \red{Observatorio del Roque de Los Muchachos, La Palma, Spain,} \citep{Osborn2015} and more recently at ESO Paranal. There is also a version under development for Mount Stromlo, Australia \citep{Grosse16}.

The wind velocity profiles from Stereo-SCIDAR have been validated with both balloon borne radiosonde and General Circulation numerical weather forecast models from the Global Forecast System \citep{Osborn16c}. This multi-way comparison shows that the numerical models are capable of forecasting wind profiles for astronomical instrumentation optimisation on average but if high-temporal resolution variations are of interest then the optical monitor is still required.

Recent applications of Stereo-SCIDAR include supporting the AO testbed, Canary \citep{Morris2014}, where it was used to validate the tomographic reconstructor as well as to validate the Linear Quadratic Gaussian smart AO controller \citep{Sivo2014b}. 

Here we specifically discuss the Stereo-SCIDAR commissioned by ESO to operate at Cerro Paranal, Chile, the site of the Very Large Telescope (VLT). The VLT comprises of four 8~m unit telescopes (UT) and four 1.8~m auxiliary telescopes (AT), and only 20~km from the site of the future ELT. Stereo-SCIDAR was commissioned at ESO Paranal in April 2016 and has been in regular operation since this date. The Stereo-SCIDAR profiles are of particular and current interest at Paranal due to the development of wide-field of view AO system on the VLT and the planned AO systems on the ELT. The optical turbulence profiles will enable performance estimation as well as performance validation of these complicated tomographic AO systems. 

We compare the results from Stereo-SCIDAR with the existing Paranal suite of dedicated site monitors, including the Surface-Layer Slope Detection And Ranging (SLODAR) instrument\citep{Wilson02,Osborn2010,Butterley2015b}, and the Multi-Aperture Scintillation Sensor - Differential Image Motion Monitor (MASS-DIMM) \citep{Sarazin1990,Sarazin2011,Kornilov03}. We also compare the wind velocity profiles from the Stereo-SCIDAR with those from the European Centre for Medium Range Weather Forecasts (ECMWF) \citep{Osborn16c}. Using the three instruments and the model in this way it is possible to validate the performance of the recently commissioned Stereo-SCIDAR instrument.

After instrumentation cross-validation we present the results, statistics and temporal variations from the Stereo-SCIDAR instrument with respect to the main applications:
\begin{itemize}
\item{
Astronomical instrumentation performance monitoring and validation. 
This will require sequences of turbulence profiles and corresponding astro-atmospheric parameters, such as seeing, coherence time and isoplanatic angle.
}
\item{
Astronomical instrument design. All future instrumentation needs to be designed for the specific atmospheric conditions they are expected to encounter. The Stereo-SCIDAR will provide distributions of astro-atmospheric parameters and the median optical turbulence profile, which is critically important for the future generation of wide-field AO.
}
\item{
Real-time instrument optimisation. 
}
\item{
General site monitoring.
}
\item{
Meso-scale atmospheric turbulence forecasting calibration and validation \citep{Masciadri2017}.
}
\end{itemize}

There has been a lot of previous work on characterising the Earth's turbulent atmosphere above ESO Paranal. 

The DIMM has been in regular operation on site since 1990 and therefore provides a large sample from which to derive integrated turbulence statistics \citep{Sarazin2008}.

Other studies such as the multi-instrument campaign of 2007 \citep{DaliAli10, Sarazin1990, Ziad04,Ramio2008, Kornilov03,Tokovinin2010, Maire2007} and the surface layer characterisation campaign of 2010 \citep{Lombardi10,Sarazin1990,Kornilov03,Osborn2010,
Tokovinin2010}, were extremely useful for understanding the atmosphere and the various instruments. However due to the limited nature of the campaigns they do not attempt to present a statistical representation of the site.

Cute-SCIDAR, another SCIDAR instrument was operational at ESO Paranal during November / December 2007 \citep{Ramio2008, Masciadri2011}. This SCIDAR operated for 20 nights and was used as part of the Paranal 2007 multi-instrument campaign \citep{DaliAli10}. The data from this instrument has proved extremely useful to further the understanding of the behaviour of the MASS \citep{Masciadri2014,Lombardi2016}.

With the exception of the MASS-DIMM these campaigns provide a limited dataset with which to compare our data. Here, we present the first 20 months of Stereo-SCIDAR operation which significantly increases the volume of high-altitude resolution and high-sensitivity turbulence profiles at ESO Paranal.

Section~\ref{sect:SCIDAR} describes the instrument, the data analysis pipeline and the data archive.
Section~\ref{sect:parameters} shows the distribution of the turbulence statistics as measured by the first phase of the Stereo-SCIDAR operation at Paranal. The comparisons of the parameters as estimated by Stereo-SCIDAR are compared to other existing instrumentation in section~\ref{sect:comparisons}. The conclusions are in section~\ref{sect:conclusions}.

\section{Stereo-SCIDAR}
\label{sect:SCIDAR}
The Stereo-SCIDAR method has been described in detail several times before (see for example \citealp{Shepherd13, Osborn2015, Osborn16c, Derie2016}).Briefly, the Stereo-SCIDAR uses the triangulation technique by cross-correlating the spatial intensity pattern (scintillation) from two stars. The offset of the correlation peak indicates the altitude of the turbulence and the magnitude of the correlation peak indicates the strength of the turbulence. The wind velocity can be estimated by measuring the velocity of the correlation peak when temporal delays are added between the images from the two stars. The advantage of Stereo-SCIDAR over previous generalised SCIDAR instruments comes from using two detectors, one for each target. This increases the sensitivity as the scintillation patterns are optically separated, rather than overlapping on a single detector which reduces the contrast. Using two detectors also enables a greater magnitude difference in the target stars, increasing the usable target catalogue and hence sky coverage.

\subsection{Data analysis pipeline}
The data analysis generally follows the routine described in \cite{Shepherd13} and \cite{Osborn16c}, with some significant changes. In \cite{Shepherd13} we fit a response function to every pixel separation. This implied a vertical resolution equal to one pixel offset, ie $z_\mathrm{max}/n_\mathrm{pix}$, where $n_\mathrm{pix}$ is the number of pixels across the pupil image of the telescope and $z_\mathrm{max}$ is the maximum propagation distance that the Stereo-SCIDAR can sense, given by $D/\theta$, where $D$ is the telescope diameter and $\theta$ is the stellar separation. However, in reality the vertical resolution of the Stereo-SCIDAR is altitude dependent, with larger propagation distances (higher turbulent zones) resulting in broader response functions (up to several pixels in size). Using a response function for every pixel separation will lead to the inverse problem being ill-conditioned. Instead, the response functions are separated by a distance of $0.5\sqrt{\lambda z}$, where $\lambda$ is the wavelength of the light and $z$ is the propagation distance. In this way the response function separation is altitude dependent and reflects the native resolution of the Stereo-SCIDAR instrument \citep{Avila98}, reducing noise due to the ill-conditioned inversion problem. We calculate the response functions for a monochromatic wavelength of 500~nm. A dichroic filter is used to reflect light with wavelengths longer then 600~nm to the acquisition camera and the Andor Luca EMCCD detectors have a cut-off at 400~nm.

The wind velocities are found by measuring the motion of the cross-covariance peaks in the temporal spatio-cross covariance function. Initially the correlation peaks are found by applying a CLEAN-like algorithm to the spatio-temporal cross-covariance function, similar to that described in \cite{Prieur04}. The velocities are then estimated by finding sets of at least five covariance peaks which appear to move in a straight line with constant velocity \citep{Osborn16c}. However, some layers can be missed in the wind velocity profile. For weak turbulence it is difficult to identify wind vectors in the Stereo-SCIDAR data and due to this limitation we can not guarantee to measure all of the turbulence velocity vectors.\comment{ The current solution is to implement a linear grid fit interpolation of all of the detected wind vectors to fill in the gaps of the measurements.}

The contribution of optical turbulence in the dome is subtracted from all of the Stereo-SCIDAR measurements automatically. Using the assumption that the dome turbulence evolves slowly we can monitor the decorrelation of the covariance peak corresponding to local turbulence and extrapolate to estimate the magnitude of the dome turbulence. This is an extension of the method proposed by \cite{Avila98} and is described in \cite{Shepherd13}.

\begin{figure*}
\centering
    \includegraphics[width=1\textwidth]{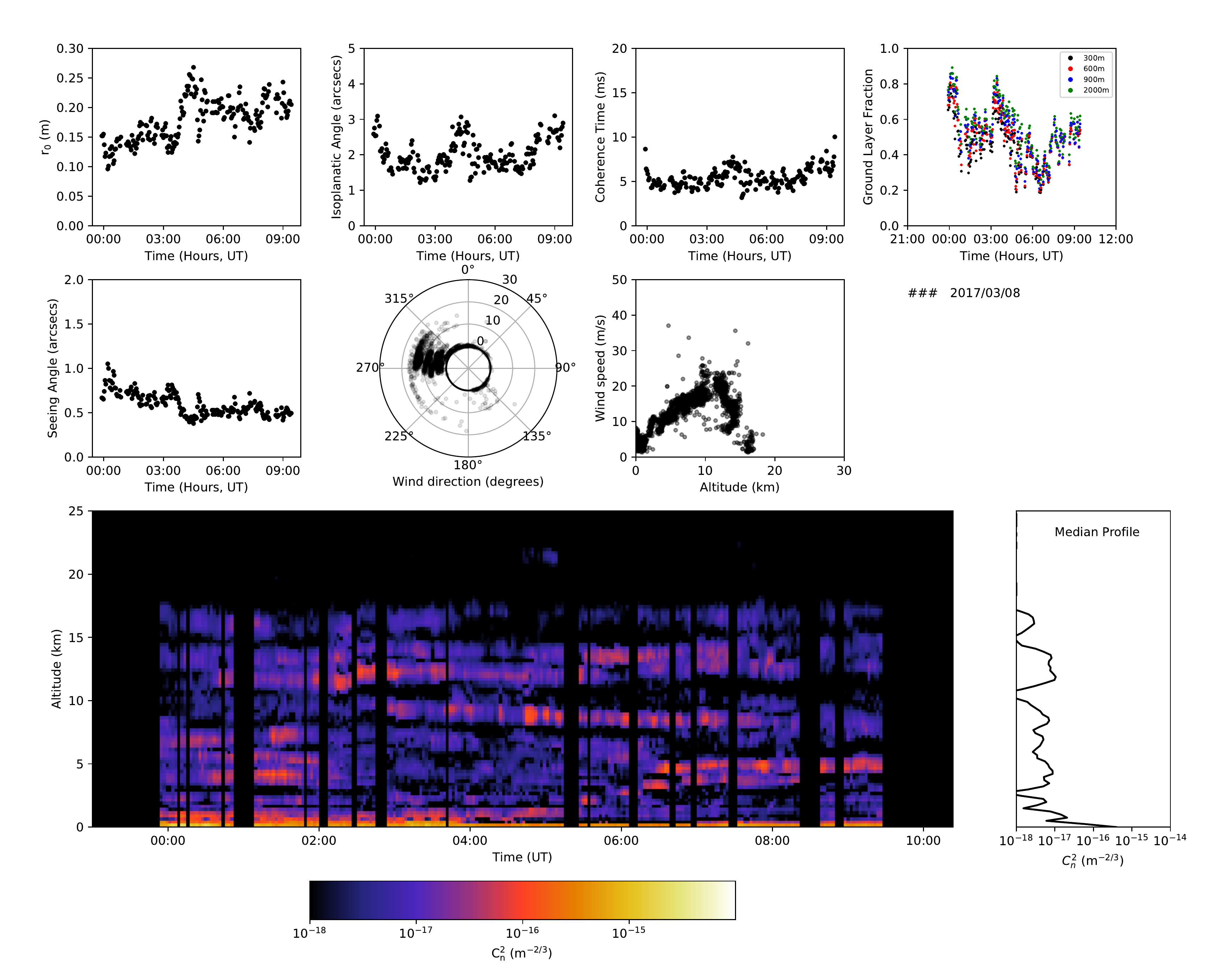}
\caption{Example Stereo-SCIDAR real-time display for the night beginning 8th March 2017. The radial numbers on the wind direction polar plot denotes the altitude above the observatory in km. The transparency of the data points denotes the time since the data was taken.}
\label{fig:20170308}
\end{figure*}

The data is analysed automatically in real time providing a real-time display which updates with new data approximately every 120\,s. Figure~\ref{fig:20170308} is an example of this real-time display.

\comment{
\subsection{Data filtering and noise}
\textbf{<?>}
}

\subsection{Data archive}
Table~\ref{tab:scidarDataParanal} summarises the data set used in this study. The total hours are found by adding together the duration of each dataset (ie it does not include gaps in the data due to change of targets or bad weather). Although the data represents times distributed throughout the year over a period of almost two years it is still a limited data set. Stereo-SCIDAR will continue to operate while it can be supported by ESO, however, here we show the results for the first phase of the project. The data is available upon request to the author and the archive will continue to grow as more data as collected. The data will be released in batches comprising a data release. The data used for this study is data release `2018A'. The profiles have been linearly interpolated into 250~m altitude bins and the normalised such as the integrated turbulence strength is conserved. The profiles are padded with `-1' above the maximum profiling altitude to maintain the same number of bins per profile. The dome seeing has been subtracted. The `native' resolution (un-interpolated) profiles are also available on request to the author.

\begin{table*}
\caption{ESO Paranal, Stereo-SCIDAR data volume: 2018A}
\label{tab:scidarDataParanal}
\begin{tabular}{@{}clccc}
\hline
Year & Month & Days & Hours & Number of Profiles\\
\hline
2016 & April    & 26 - 29	& 18.43	& 607\\
	 & July		& 22 - 26	& 37.12	& 1143\\
     & October  & 30 - 31	& 10.65	& 301\\
     & November & 1 - 2		& 10.80	& 302\\
     & December & 10 - 12	& 11.62	& 308\\
2017 & March	& 7 - 9		& 16.46	& 469\\
	 & April	& 12 - 18	& 37.34	& 988\\
     & May		& 5 - 9		& 16.06	& 419\\
     & June		& 8 - 10	& 19.97	& 511\\
     & July		& 3 - 9		& 37.60	& 962\\
     & August	& 3 - 8		& 34.42 & 930\\
     & November & 4 - 9, 18 - 20, 29 - 30 & 45.63 & 1076\\
     & December & 1 - 6, 8 - 18 & 56.69 & 1483 \\
2018 & January	& 13 - 24   & 44.19	&	1192\\
\hline
Totals:	&		& 83	&	396.97 & 10691	\\
\hline
\end{tabular}
\end{table*}

\section{Parameter statistics}
\label{sect:parameters}
\begin{figure}
\centering
    \includegraphics[width=0.5\textwidth,]{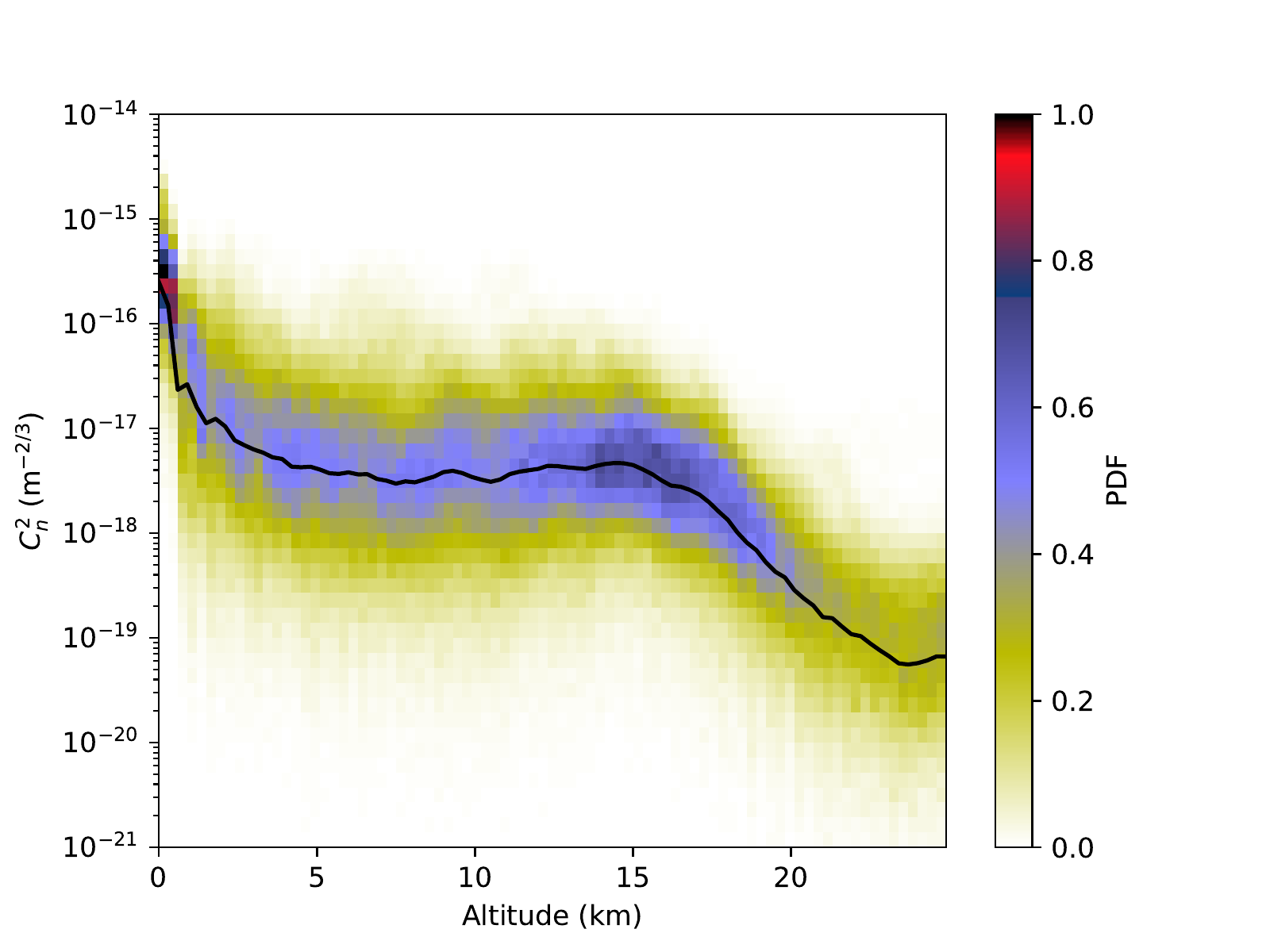}
\caption{The median optical turbulence strength profile from Stereo-SCIDAR at the Cerro Paranal. The black line shows the median. The colour shows the distribution of the turbulence strength at each altitude.}
\label{fig:scidarMedian}
\end{figure}

Figure~\ref{fig:scidarMedian} shows the median turbulence strength as a function of altitude above observatory level. The median profile can not be used as a typical profile. However, it does give an estimate of the expected turbulence strength at each altitude.

Stereo-SCIDAR provides measurements of the strength of the optical turbulence, quantified by the refractive index structure parameter, $C_n^2(h)$, as a function of altitude, $h$,  and the turbulence speed, $|V(h)|$, and direction $V_\theta(h)$. Using these functions it is possible to derive a number of other optical parameters:
\begin{equation}
r_0 = \left(0.423 \left(\frac{2\pi}{\lambda}\right)^2 \cos(\gamma)^{-1} \int C_n^2(h) \mathrm{d}h \right)^{-3/5},
\end{equation}
\begin{equation}
\epsilon = \frac{0.98\lambda}{r_0},
\end{equation}
\begin{equation}
\theta_0 = \left(2.914 \left(\frac{2\pi}{\lambda}\right)^2 \cos(\gamma)^{-8/3} \int C_n^2(h) h^{5/3} \mathrm{d}h \right)^{-3/5},
\end{equation}
\begin{equation}
\tau_0 = \left(2.914 \left(\frac{2\pi}{\lambda}\right)^2 \cos(\gamma)^{-8/3} \int \frac{C_n^2(h)}{V(h)^{5/3}} \mathrm{d}h \right)^{-3/5},
\end{equation}
\begin{equation}
\label{eqn:scint_index}
\sigma^2_I = 10.7D^{-4/3}t^{-1}\cos(\gamma)^{\alpha (V_\theta (h))}\int \frac{C_n^2(h) h^2}{V(h)} \mathrm{d}h,
\end{equation}
where $r_0$ is the Fried parameter \citep{Fried1966}, $\epsilon$ is the Full Width at Half Maximum (FWHM) of the point spread function (PSF) or seeing on a small telescope, $\theta_0$ is the isoplanatic angle \citep{Roddier1981}, $\tau_0$ is the coherence time \cite{Greenwood78}, $\sigma^2_I$ is the scintillation variance on a telescope of diameter larger than a few tens of centimetres and for long exposures \citep{Dravins1997,Osborn2015a}. Other required parameters for the above calculations are the zenith angle, $\gamma$, the wavelength of the observation, $\lambda$, the telescope diameter, $D$, \red{the observation exposure time, $t$}, and the air mass exponent, $\alpha$. Note that the value of the airmass exponent, $\alpha$, will depend on the wind direction and vary between -3 for the case when the wind is transverse to the azimuthal angle of the star, up to -4 in the case of a longitudinal wind direction. This is a geometric correction. In the case where the wind direction is parallel to the azimuthal angle of the star, the projection of the telescope pupil onto a horizontal layer is stretched by a factor of $1/\cos(\gamma)$, which changes the projected wind speed. Therefore, $\alpha = -3.5+\cos(2(V_\theta - \theta_{\mathrm{az}}))/2$, where $\theta_{\mathrm{az}}$ is the azimuthal angle of the observation. Further discussion of scintillation in astronomical time-resolved photometry for smaller telescopes and short exposures can be found in \cite{Osborn2015a}. 

Each of these parameters has its own influence for particular applications. $r_0$ and $\epsilon$ are both used to measure of the effect on an image caused by a wavefront which has propagated through the complete atmosphere. The isoplanatic angle defines the angular extent over which the atmospheric effects are correlated. It is this parameter that defines the angular size of an Adaptive Optics (AO) corrected field. Multi-Conjugate Adaptive Optics (MCAO) systems can be used to increase the isoplanatic angle and hence increase the corrected field of view. The coherence time defines the update rate that an AO system must function at in order to minimise residual wavefront errors due to the temporal lag between the wavefront measurements and correction by the deformable mirror (DM). All of this information can be used in real time for AO support, PSF reconstruction, observatory and observation scheduling.

The scintillation index is critical for time resolved photometry. Here, we show the scintillation index for a 1~m telescope and 1~s exposure, such that it can easily be scaled to other system specifications.

The fraction of the turbulence in the ground layer is also a parameter of significant interest to observatories with interests in wide-field of view AO instrumentation, such as Ground-Layer AO. The performance and the uniformity of correction of wide-field AO systems is very dependent on the structure of the atmospheric optical turbulence profile. Here we present the ground layer fraction \red{,defined as the ratio of the turbulence strength up to the given altitude to the integrated turbulence up to the maximum sensing altitude,} (GF) up to 300, 600, 900 and 1200~m.

In addition to the above, statistical data on the typical profiles and variability of each of the profiles can be used for instrument development and performance analysis \citep{Morris14}.

In table~\ref{tab:scidarPar} we show the first, second and third quartile of each of the parameters of interest. 

It is interesting to compare these values to previous studies. In \cite{Sarazin2008} the authors show that the seeing at Paranal, as measured by the original DIMM has actually increased over the years from a median of 0.65\as  to a median of 1.1\as, whereas the seeing from the UT image quality measurements of FORS2, an instrument on the VLT, has remained at 0.65\as. The authors of that study suggest that this discrepancy is likely caused by a thin strong ground layer which is becoming more frequent and the \red{effect on the original DIMM was exacerbated by its location close to the 20~m high VLT Survey Telescope. The instruments on the UTs are protected from this low altitude turbulence by the telescope dome. In this study we compare with a new DIMM in a new location, further from any buildings and on a higher tower.} The new DIMM reports a median seeing of 0.63\as (during Stereo-SCIDAR runs), which is compatible with the Stereo-SCIDAR measurements and indeed with the UT image quality measurements.

\begin{table}
\caption{Astro-atmospheric parameter statistics for Paranal for the Stereo-SCIDAR from data release 2018A.}
\label{tab:scidarPar}
\begin{tabular}{@{}lccc}
\hline
Parameter & Q1 & Median & Q3 \\
\hline
Seeing 					& 0.52\as	& 0.64\as	& 0.85\as\\
Coherence Time			& 2.82 ms	& 4.18 ms	& 6.65 ms	\\
Isoplanatic Angle		& 1.34\as	& 1.75\as	& 2.22\as	\\
Scintillation Index	(1m,1s,$\times10^{-5}$)	& 0.39	& 0.63		& 1.04 		\\
GF ($h$<300 m)			& 0.14		& 0.25		& 0.38		\\
GF ($h$<600 m)			& 0.25		& 0.40		& 0.57		\\
GF ($h$<900 m)			& 0.29		& 0.45		& 0.62		\\
GF ($h$<1200 m)			& 0.34		& 0.49		& 0.65		\\
\hline
\end{tabular}
\end{table}

\begin{figure}
\centering
    \includegraphics[width=0.5\textwidth,trim={0.5cm 0cm 0cm 0cm},clip]{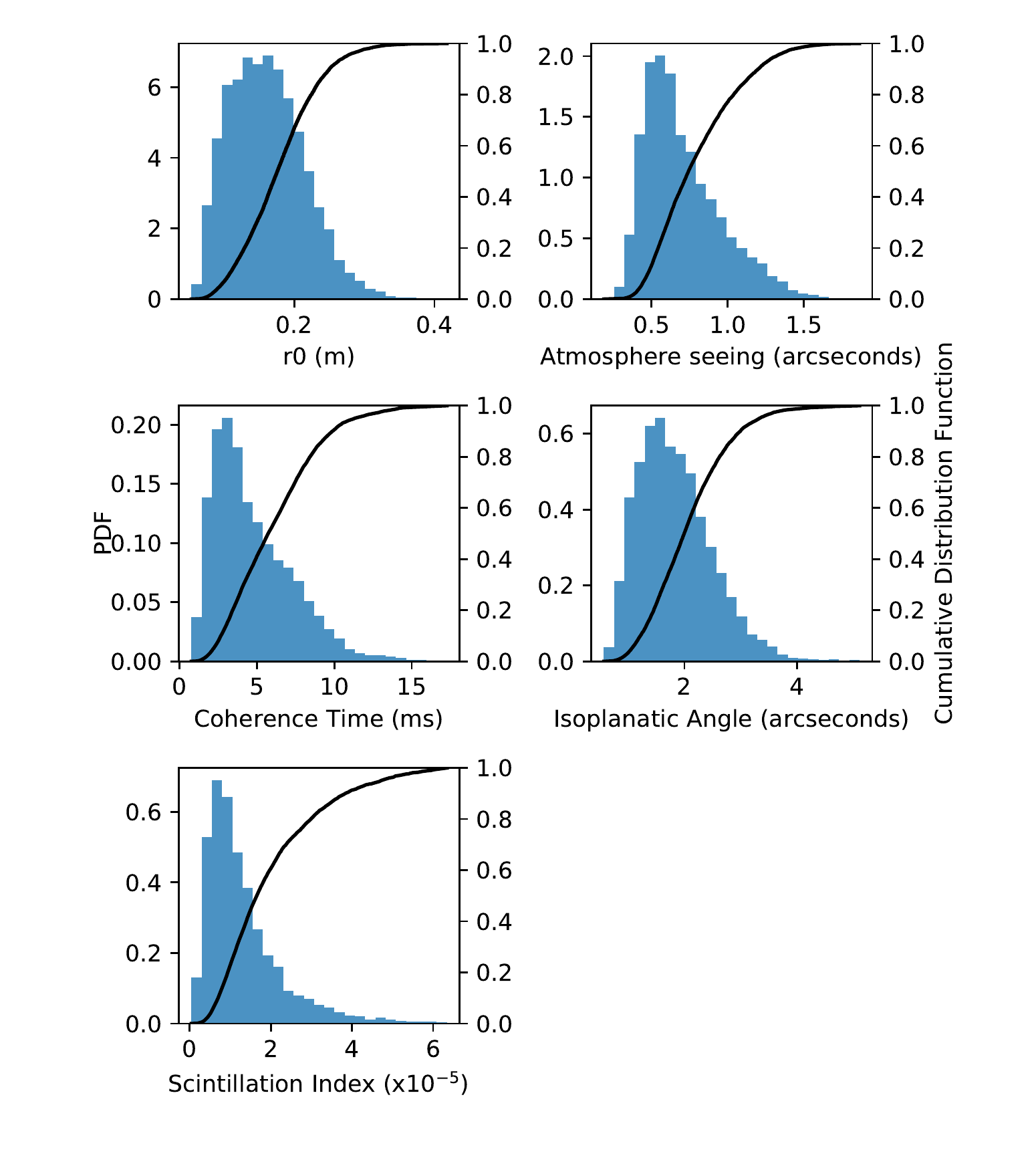}
\caption{The distribution of $r_0$, seeing, coherence time, isoplanatic angle and scintillation noise (scaled to 1~m telescope with 1~s exposure time) with cumulative density function overlaid as solid line.}
\label{fig:scidarParameters}
\end{figure}

\begin{figure}
\centering
    \includegraphics[width=0.5\textwidth,trim={0.5cm 1cm 0cm 0.5cm},clip]{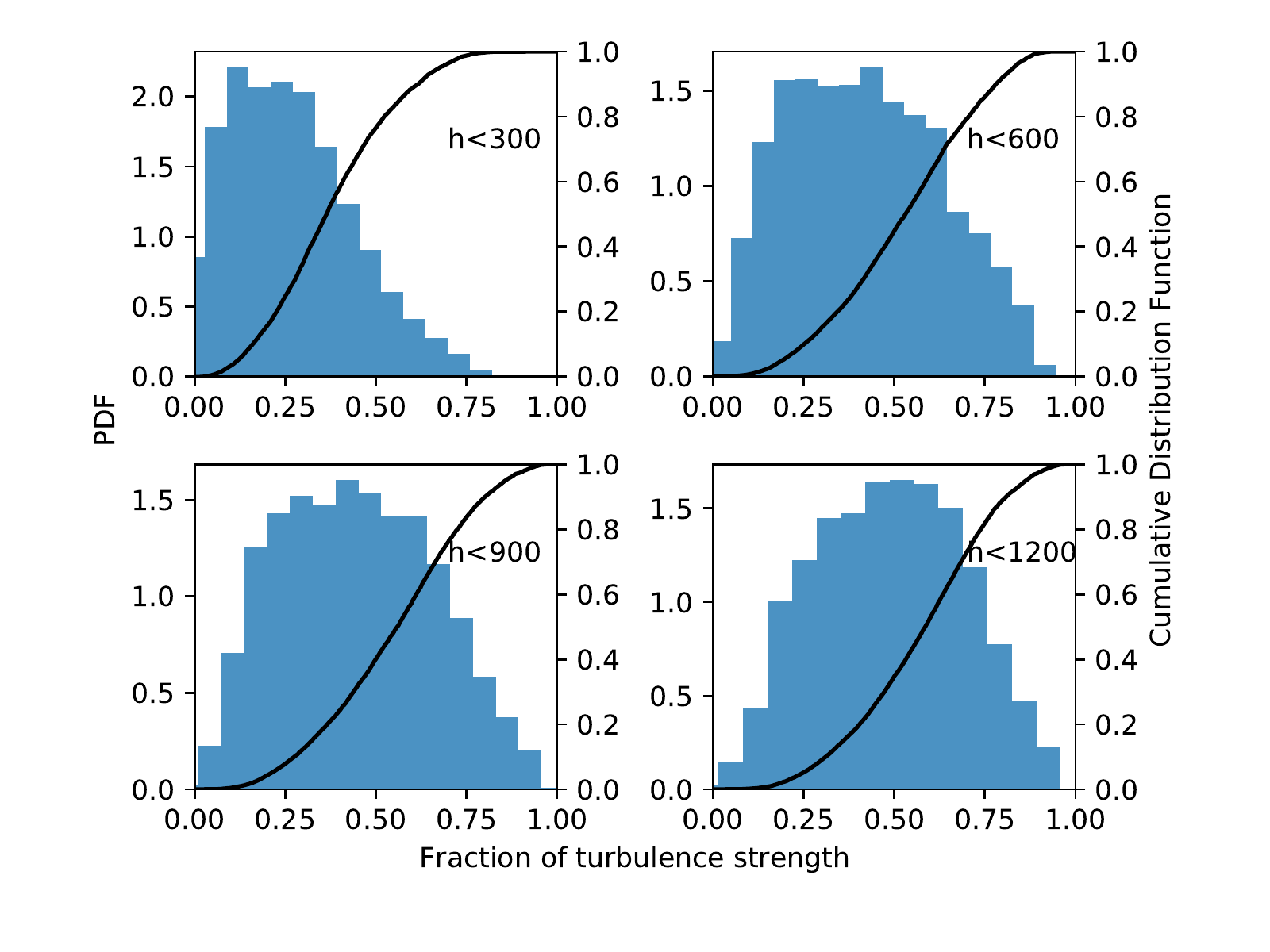}
\caption{Distributions of the fractions of turbulence below an altitude $h$. The fractional turbulence distributions, together with the cumulative distribution, is shown for four altitudes, $h<$ 300~m, 600~m, 900~m and 1200~m (top left, top right, bottom left, bottom right respectively). The fraction of turbulence can be seen increasing as we integrate up to higher altitudes, as expected. However, the width of the distribution is of particular interest. On some occasions, up to 75~\% of the turbulence can be found in the first 300~m.}
\label{fig:GF}
\end{figure}

\begin{figure}
\centering
    \includegraphics[width=0.5\textwidth,trim={0cm 0cm 0cm 1cm},clip]{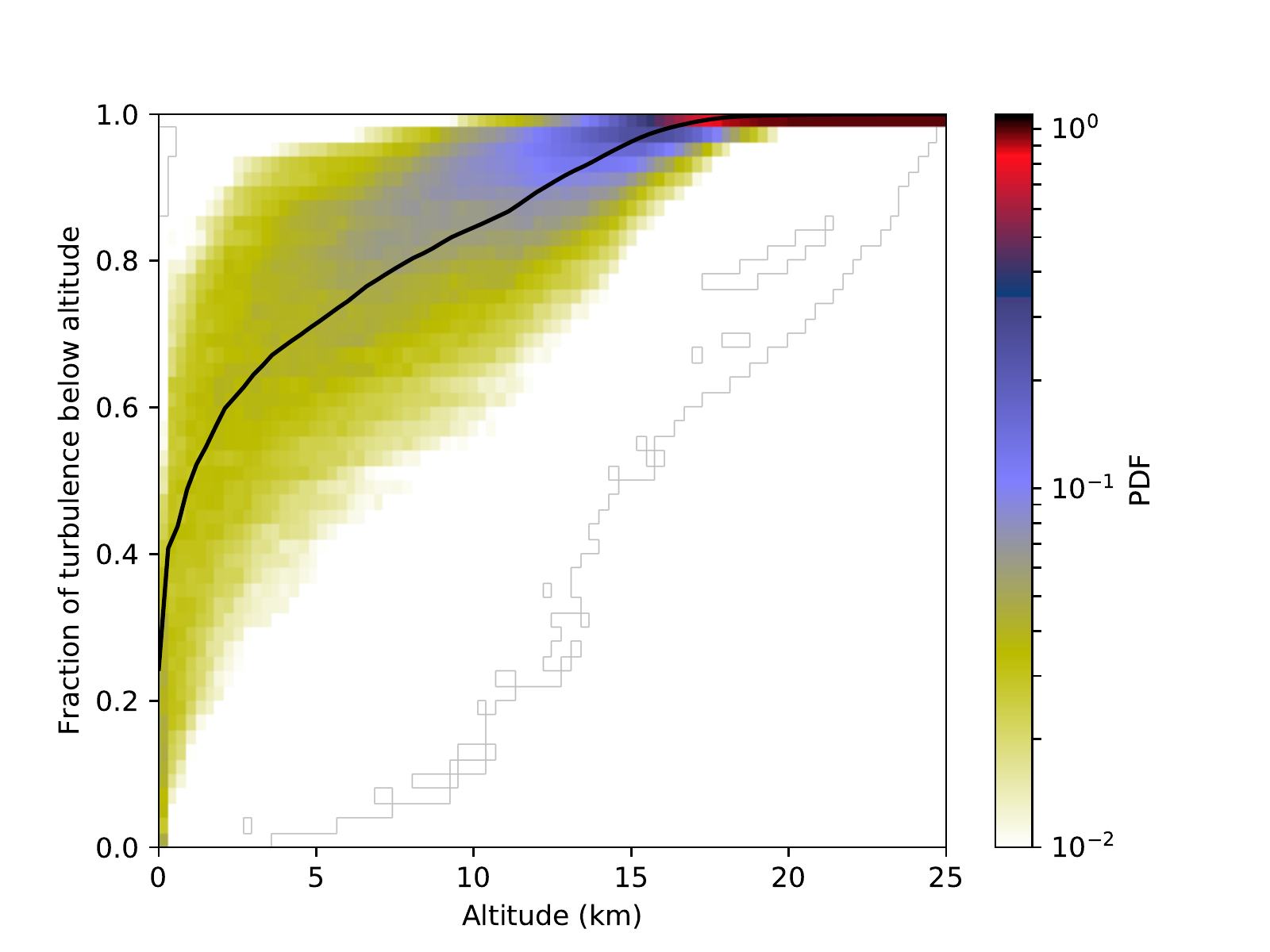}
\caption{Distribution of integrated turbulence up to an altitude. The colour indicates the distribution of the fraction of turbulence below a given height, such that a wide spread indicates a large range of values. The median is shown by the solid line. For example, approximately 50\% of the turbulence strength is confined to an altitude below 2.0~km, however this value can actually range between 0.2 and 0.8 as shown by the colour.}
\label{fig:histAltDist}
\end{figure}

The distributions of these parameters are shown in Fig.~\ref{fig:scidarParameters} and Fig.~\ref{fig:GF}. Fig.~\ref{fig:histAltDist} shows the distribution of the fraction of the turbulence up to any given altitude. For example, the figure shows that approximately 50\% of the turbulence strength is confined to an altitude below 2.0~km, however this value is variable and can actually range between 0.2 and 0.8.

\section{Instrument Comparisons}
\label{sect:comparisons}
The Stereo-SCIDAR outputs, turbulence strength and velocity vertical profiles, are compared with other instruments on the Paranal site. Comparisons are made for measurements between the Stereo-SCIDAR and the alternative instrument / model for measurements taken within 5 minutes of each other. If more than one measurement was made within the time frame the median value is used in the comparison. \red{The mean value provides a very similar answer to the median (\textless 1\% difference in comparison parameters), and so is not reported here.} The instruments were spatially separated and were not observing the same targets. The parameters for all instruments are corrected for zenith angle.

Table~\ref{tab:scidarComp} shows the instruments and their corresponding metrics for the comparisons. We see that generally the correlation between the instrumentation (and the model in the case of the ECMWF) is high. In the following sections we will discuss each comparison in more detail. We also show the values for the Stereo-SCIDAR compared with itself averaged over the 5 minute comparison period. This is done to show the spread of the data expected from comparing two measurements in the sampling period. 

\begin{table*}
\caption{Astro-atmospheric parameter comparison for Paranal for the Stereo-SCIDAR with the Surface Layer SLODAR, MASS-DIMM and the wind velocities from the ECMWF. The comparison is in terms of the Pearson correlation coefficient (C), bias and root-mean-square error (RMSE). We also show the values for the Stereo-SCIDAR compared with itself averaged over the 5 minute comparison period.}
\label{tab:scidarComp}
\begin{tabular}{@{}llccc}
\hline
Instrument & Parameter & C & Bias & RMSE\\
\hline
SL-SLODAR       & Seeing (h>5 m)	& 0.73	& -0.04\as	& 0.20\as\\
MASS-DIMM		& Seeing			& 0.84	& -0.11\as 	& 0.21\as\\
				& Free Atmosphere Seeing (MASS)		& 0.84	& 0.02\as 	& 0.14\as\\
 				& Coherence Time	& 0.73 	& -0.11ms 	& 2.01ms\\
 				& Isoplanatic Angle	& 0.61	& -0.10\as 	& 0.63\as \\
ECMWF			& Wind Speed		& 0.82	& 0.41 m/s	& 6.44 m/s\\
				& Wind Direction	& 0.77	& 1.59 deg	& 27.09 deg\\
Stereo-SCIDAR	& Seeing			& 1.00	& 0.06\as	& 0.24\as	\\
				& Coherence time	& 1.00	& 0.51ms	& 2.20ms	\\
                & Isoplanatic angle	& 1.00	& 0.12\as	& 0.69\as	\\
\hline
\end{tabular}
\end{table*}

\begin{figure*}
\centering
	\includegraphics[width=0.95\textwidth]{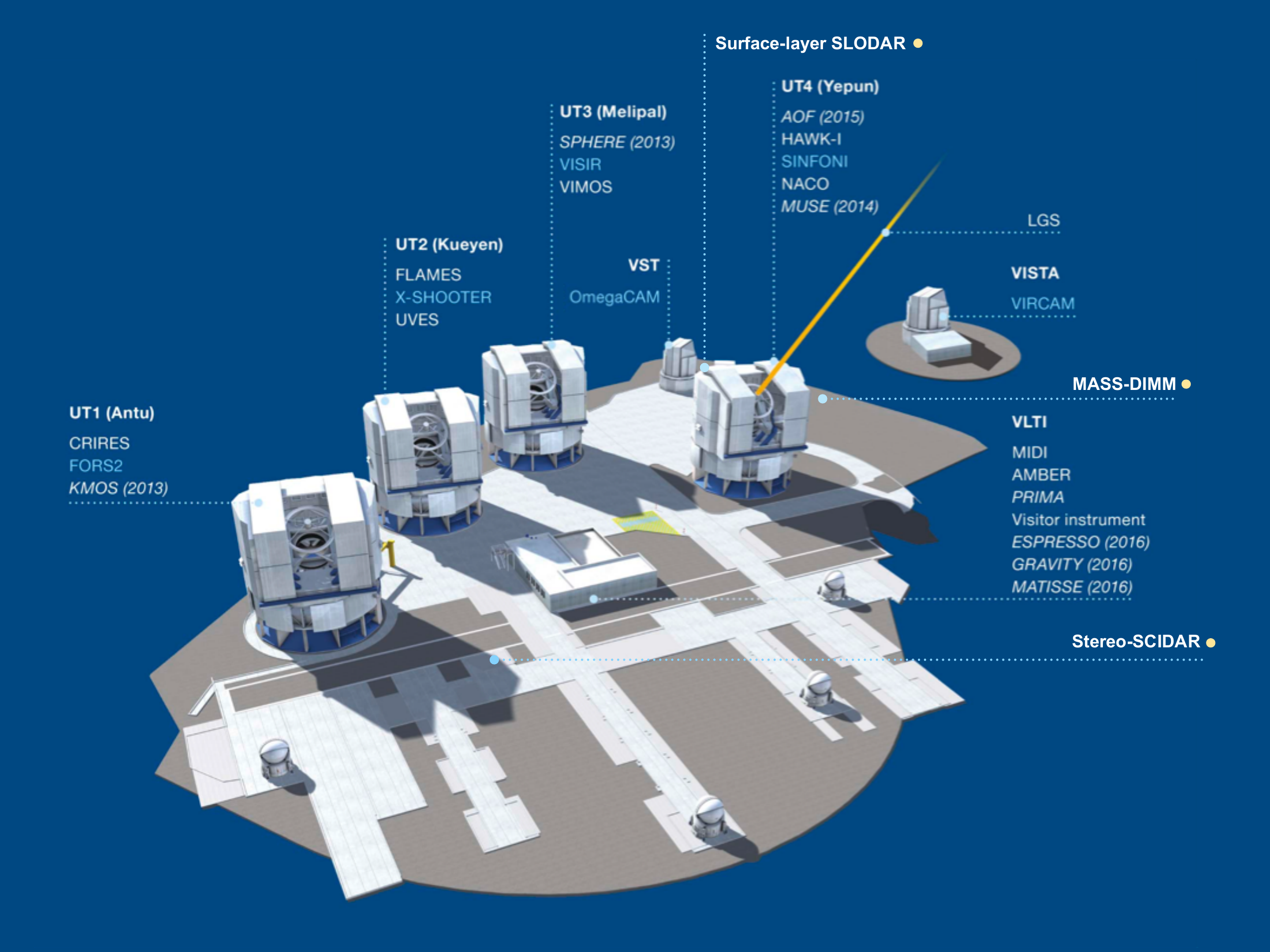}
\caption{Overview of the instrumentation at the ESO Paranal site. Modified from original image by ESO to include the locations of the turbulence monitoring instrumentation (indicated by yellow markers).}
\label{fig:VLT_map}
\end{figure*}

The map of the Paranal site complete with telescopes and instrumentation is shown in figure~\ref{fig:VLT_map}. It can be seen that the Stereo-SCIDAR is located in the centre of the platform, whereas the other site monitoring instrumentation is located near the edge of the platform. This may have an effect on the magnitude of the measured optical turbulence near the ground.

\subsection{MASS-DIMM}
The MASS-DIMM is a combination of two instruments: a DIMM to measure the integrated seeing \citep{Sarazin1990} and a MASS channel to perform low resolution profiling \citep{Tokovinin07}. The MASS-DIMM also estimates the isoplanatic angle from these low resolution profiles and the coherence time from the variance of the logarithm of the intensity ratio for different exposure times \citep{Sarazin2011}.

Due to technical issues the MASS-DIMM was unavailable between 1st February 2017 to 19th May 2017. For this reason we only have 68 nights of overlap between the MASS-DIMM and the Stereo-SCIDAR.

\begin{figure*}
\centering
$\begin{array}{cc}
	\includegraphics[width=0.5\textwidth,trim={2cm 0cm 0cm 0cm},clip]{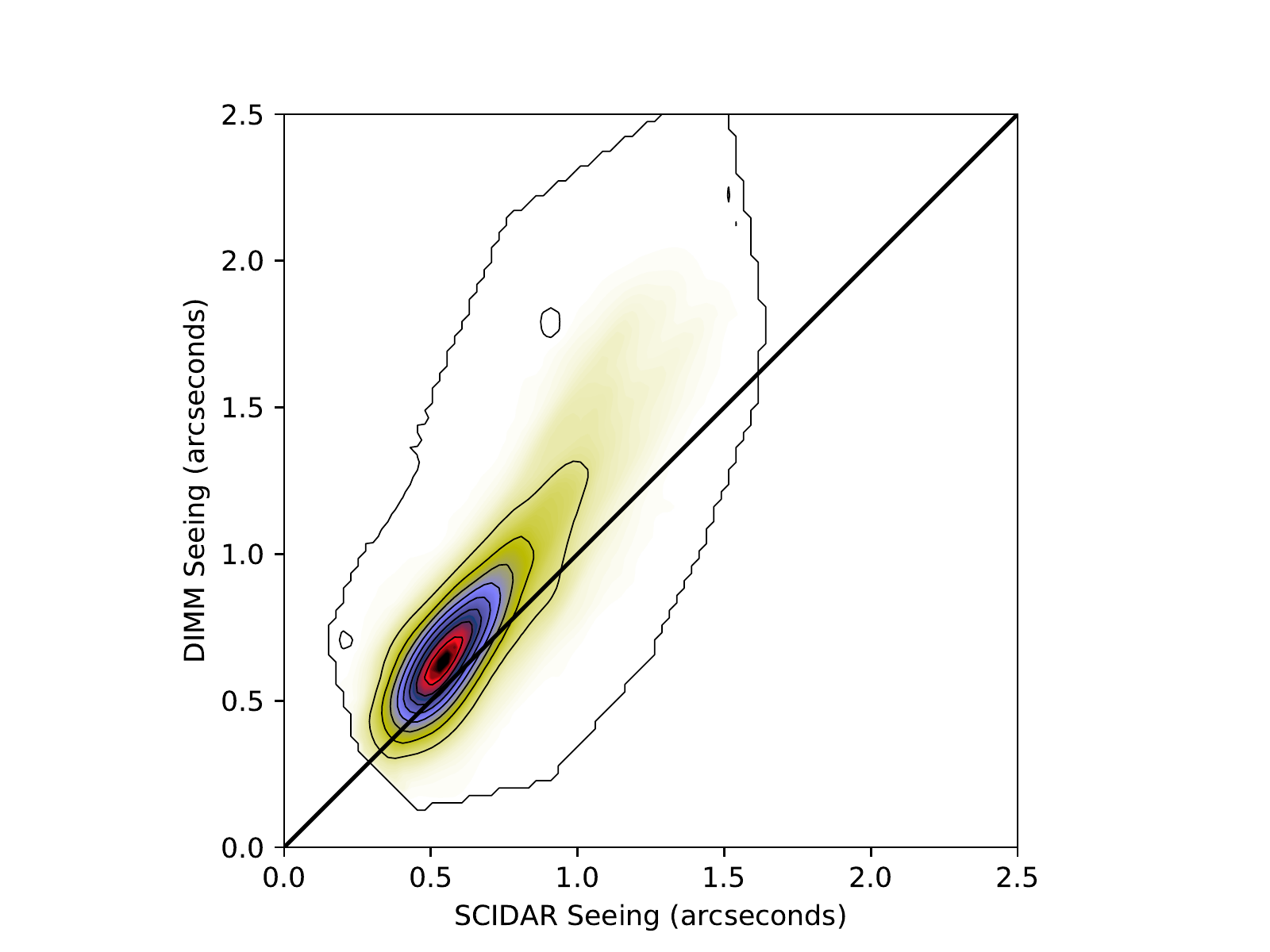}&
    \includegraphics[width=0.5\textwidth,trim={2cm 0cm 0cm 0cm},clip]{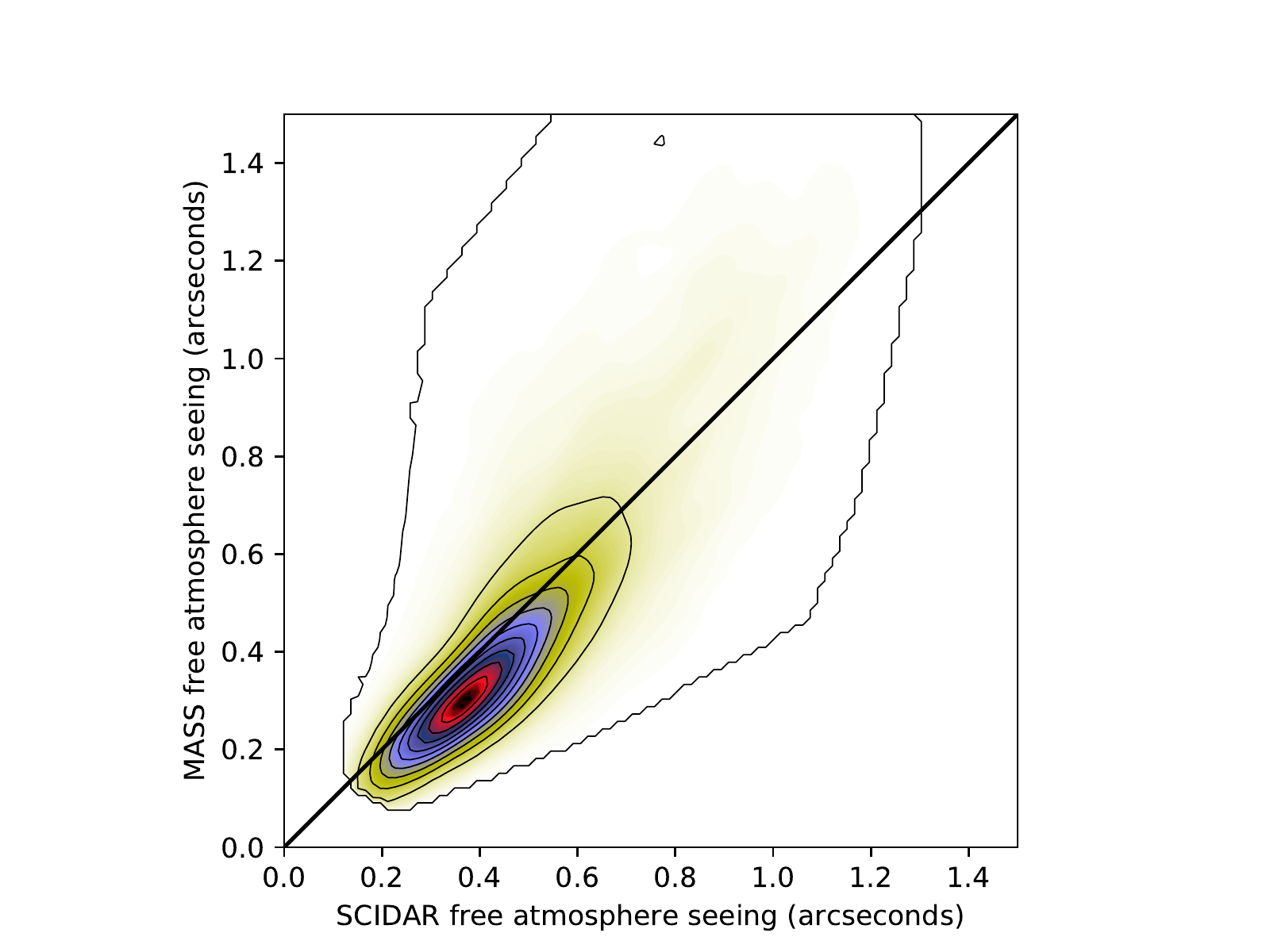}\\
    \includegraphics[width=0.5\textwidth,trim={2cm 0cm 0cm 0cm},clip]{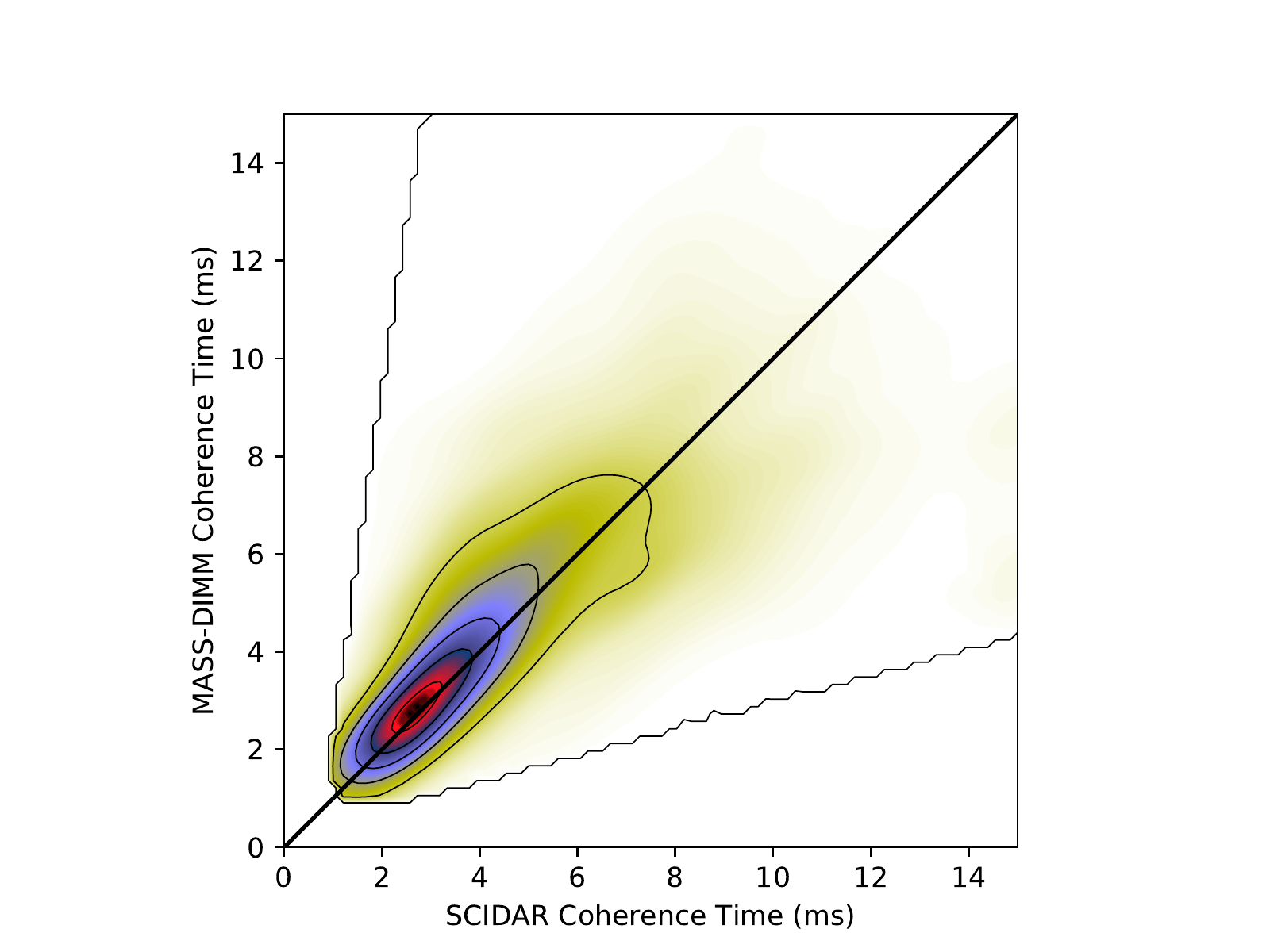}&
    \includegraphics[width=0.5\textwidth,trim={2cm 0cm 0cm 0cm},clip]{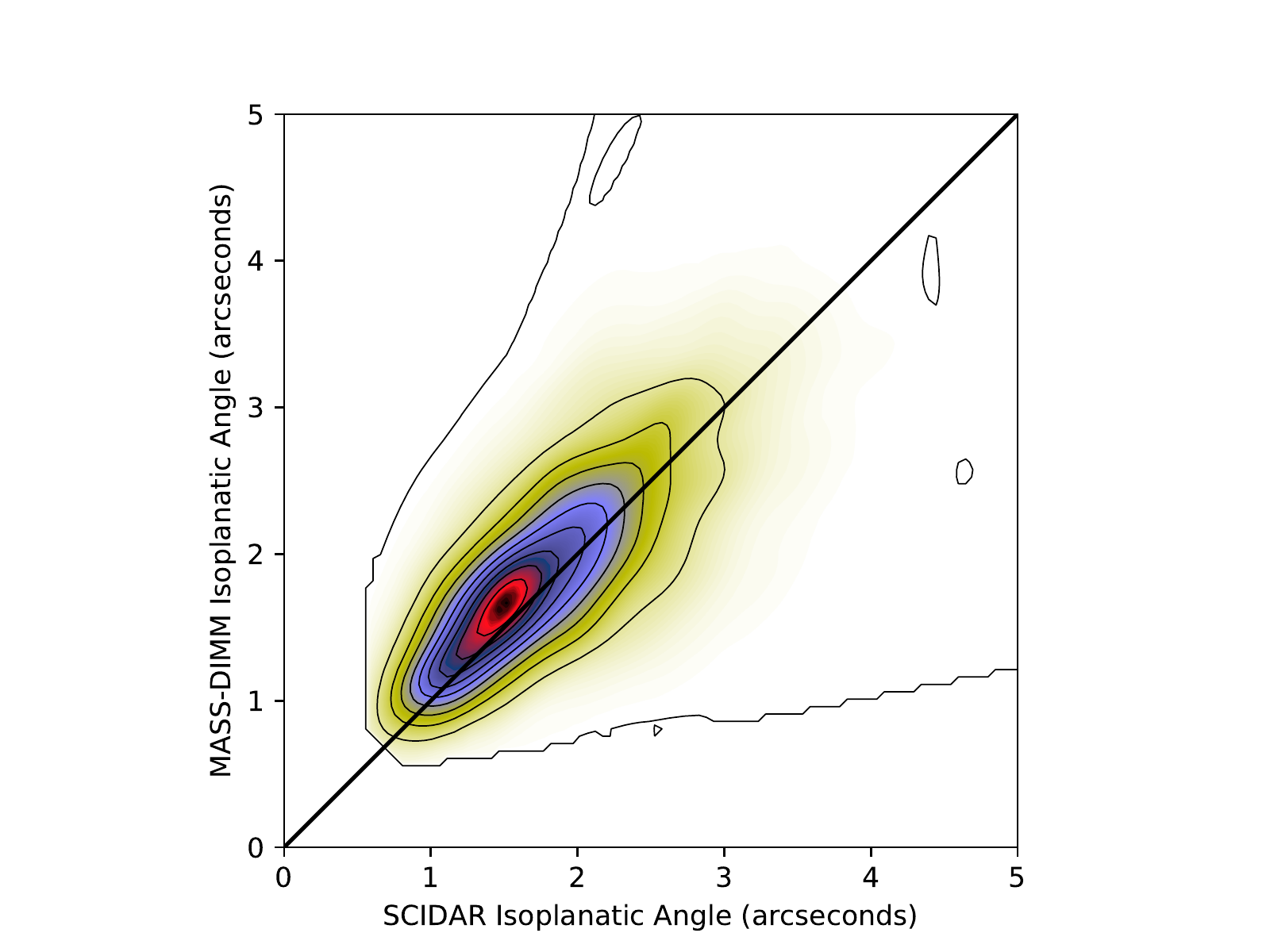}\\
\end{array}$
\caption{Astro-atmospheric parameter comparison between the MASS-DIMM and the Stereo-SCIDAR for seeing (top, left), free-atmosphere seeing (top, right), coherence time (bottom, left) and isoplanatic angle (bottom, right). The comparison metrics can be found in table~\ref{tab:scidarComp}.}
\label{fig:massScidarComparisons}
\end{figure*}

\begin{figure}
\centering
	\includegraphics[width=0.5\textwidth]{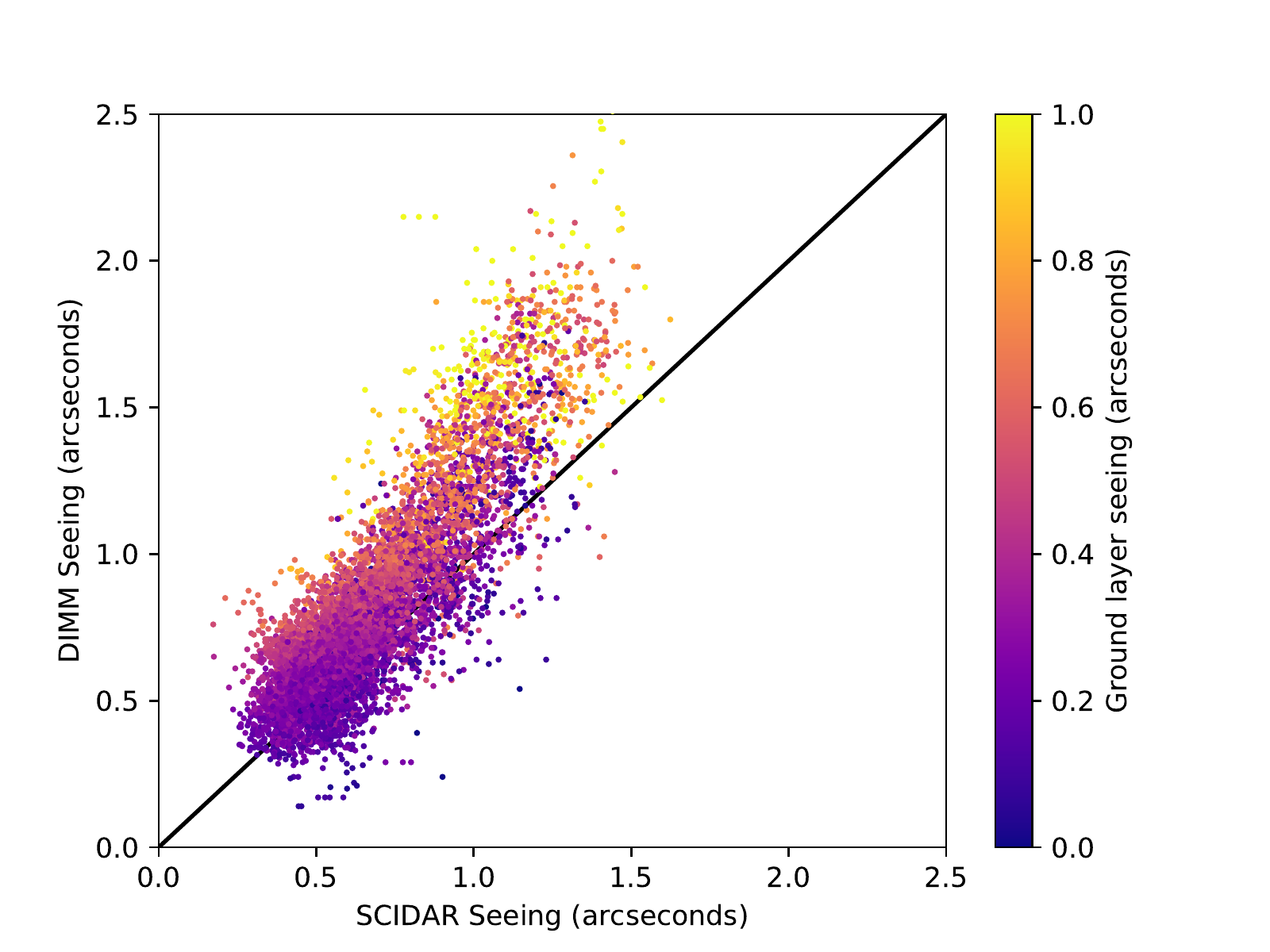}
\caption{Comparison of the Stereo-SCIDAR and DIMM seeing, coloured by the strength of the ground layer (DIMM seeing - MASS seeing).}
\label{fig:massScidarSeeingGround}
\end{figure}

Figure~\ref{fig:massScidarComparisons} shows the comparison between the Stereo-SCIDAR and the MASS-DIMM for the integrated seeing, free atmosphere seeing, the coherence time and the isoplanatic angle. In all cases the correlation is high, between 0.61 for the isoplanatic angle and 0.84 for the seeing. The free atmosphere seeing is found by projecting the Stereo-SCIDAR onto the MASS-DIMM weighting functions to take into account the non-uniform response of the MASS-DIMM to turbulence, particularly in the 250~m to 500~m altitude range. The reason for the lower correlation in the isoplanatic angle comparison might be due to the low altitude resolution profiles from the MASS used in the calculation, as suggested by the relatively large RMSE but low bias. The shape of the seeing comparison curve is interesting as it shows a trend for the Stereo-SCIDAR to measure less turbulence in stronger seeing conditions. This could either be due to a bias in one of the instruments in bad seeing (due to scintillation saturation in the Stereo-SCIDAR for example) or a physical manifestation due to the spatial separation of the instruments. The Stereo-SCIDAR is located in the centre of the observing platform, whereas the MASS-DIMM is located at the edge. Therefore the MASS-DIMM may encounter elevated seeing due to the strong turbulence at the edge of the platform in certain conditions (certain wind directions for example). From Fig.~\ref{fig:massScidarSeeingGround} we can see that the high seeing tail corresponds to periods of high ground layer strength. This supports the argument that the difference is due to location of the instruments (as suggested by \citealp{Sarazin2008}).

\subsection{Surface-Layer SLODAR}
The Surface-Layer SLODAR is a fully robotic and automatic turbulence profiler designed to profile the lowest region of the Earth's atmosphere with high vertical resolution \citep{Osborn2010,Butterley2015b}. SLODAR is a crossed-beams technique, like SCIDAR, which makes use of a Shack-Hartmann wavefront sensor to observe bright double star targets. As the method is based on direct measurements of the wavefront phase gradient, it is relatively straightforward to calibrate in terms of the absolute optical turbulence profile \citep{Butterley06}. Surface-Layer SLODAR uses wide optical binary stars to probe the lower atmosphere, up to a few hundred metres, with an altitude resolution of a few tens of metres. The exact values depend on the separation of the target stars and airmass.

\begin{figure}
\centering
    \includegraphics[width=0.5\textwidth,trim={1cm 0cm 0cm 1cm},clip]{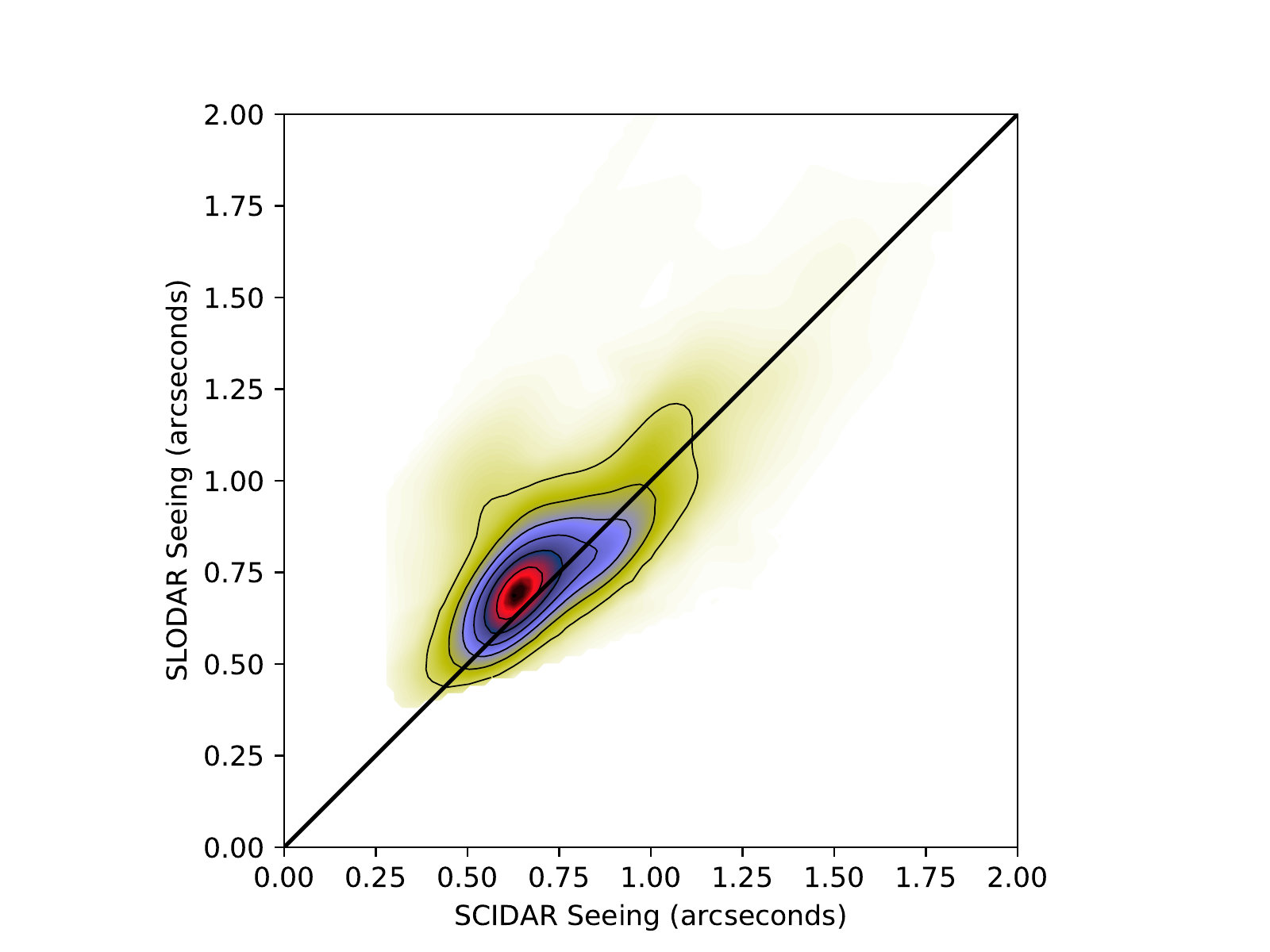}
\caption{Comparisons of Stereo-SCIDAR and SL-SLODAR for the integrated seeing from the altitude of the AT upwards. The correlation is 0.73, bias is -0.04\as and RMSE is 0.2\as.}
\label{fig:slodarScidarGS3}
\end{figure}

Fig.~\ref{fig:slodarScidarGS3} shows the comparison of the Stereo-SCIDAR and the Surface-Layer SLODAR integrated seeing from the height of the AT (5~m) upwards. We see a high correlation between the two measurements, however there is a small bias for Surface-layer SLODAR to measure more integrated turbulence than the Stereo-SCIDAR. This is likely to be due to the locations of the instruments. The Stereo-SCIDAR is located on the centre of the observing platform away from the edges of the mountain. However, the Surface-Layer SLODAR, like the MASS-DIMM, is located near to the edge of the platform where local turbulence can be higher.

\subsection{ECMWF}
General circulation models (GCM) have been used to provide wind velocity profiles for previous astronomical studies (for example, \citealp{Hagelin2010,Osborn16c}). They have also been used as the input for mesoscale turbulence forecast models (for example, for the wind velocity profile \citealp{Masciadri2013b} and for the turbulence strength profile \citealp{Giordano2013,Masciadri2017}). In this study we use the European Centre for Medium-range Weather Forecasts (ECMWF) to compare with the wind velocity measurements recovered from the Stereo-SCIDAR instrument\footnote{https://www.ecmwf.int/en/forecasts/datasets/}. 

The ECMWF model is a non-hydrostatic model. The model is refreshed every 6 hours and provides a forecast for every hour. Two level models are produced, pressure level and model level. For the model levels, as used here, forecasts are provided at 137 altitude levels. The altitude levels are hybrid, defined as lines of constant pressure above surface pressure. The altitude resolution is generally a couple of tens of metres near the ground and a few kilometres above the tropopause.

Here, we use publicly available data from ECMWF from the ERA5 catalogue. The data has 0.3~degree spatial resolution and is only available for the models produced at 06:00 and 18:00 UT, with forecasts for every hour up at 19 hours. Here, we use the best case data, i.e. data that was produced at most 11 hours before (for example, 06:00+11 hours). To extract the parameters for the site of Cerro Paranal in the 0.3~degree grid, we linearly interpolate between the four nearest data points.

\begin{figure}
\centering
    \includegraphics[width=0.5\textwidth,trim={0cm 0cm 0cm 0cm},clip]{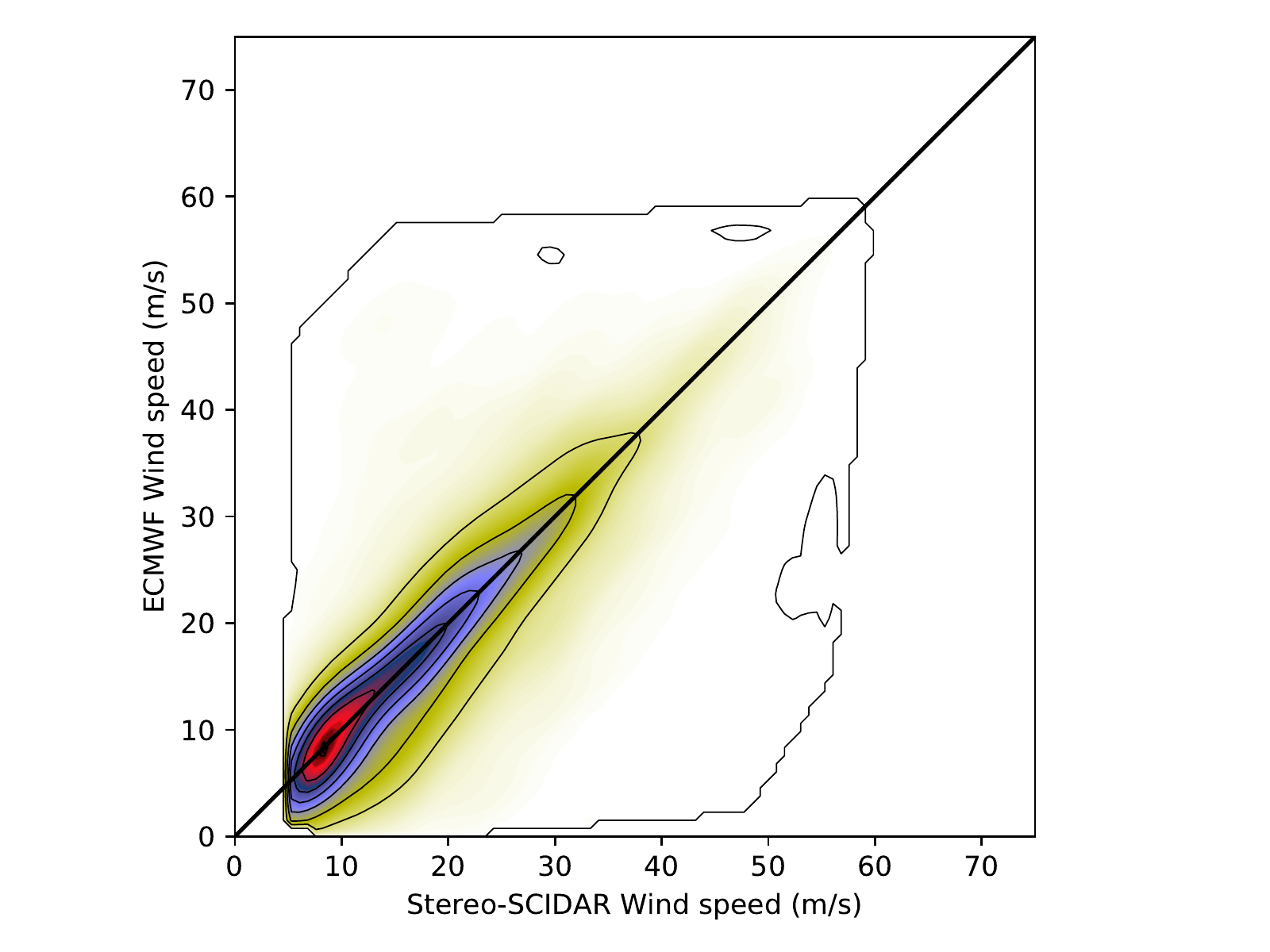}
  \caption{Comparisons of Stereo-SCIDAR recovered wind speed with the forecast wind speed from the ECMWF. The wind speed correlation = 0.82, bias = 0.41~m/s and RMSE = 6.44~m/s.}
\label{fig:ecmwfScidarSpeedComparisons}
\end{figure}

\begin{figure}
\centering
    \includegraphics[width=0.5\textwidth,trim={0cm 0 0cm 0},clip]{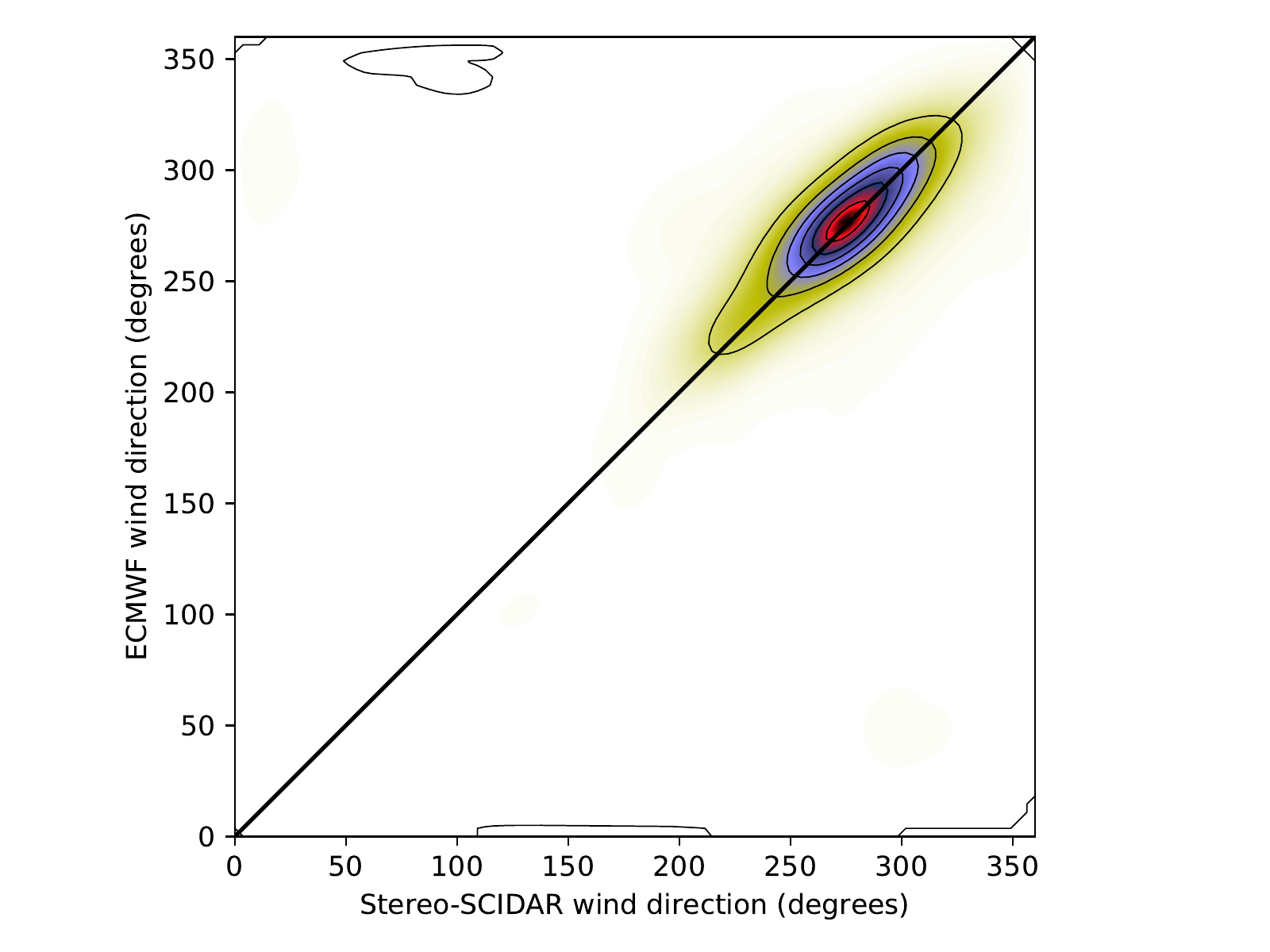}
\caption{Comparisons of Stereo-SCIDAR recovered wind direction with the forecast wind direction from the ECMWF. The wind direction correlation = 0.77, bias = 1.59~degrees and RMSE = 27.09~degrees.}
\label{fig:ecmwfScidarDirectionComparisons}
\end{figure}

Fig.~\ref{fig:ecmwfScidarSpeedComparisons} shows the comparison between wind speed from the Stereo-SCIDAR and ECMWF for all altitudes, Fig.~\ref{fig:ecmwfScidarDirectionComparisons} shows the comparison of wind direction. The correlation values of this comparison (0.82 and 0.77 for wind speed and direction respectively) are lower than those found for La Palma as reported in \cite{Osborn16c} (0.9 and 0.93 respectively). We also see that the RMSE of the wind direction is significantly larger at Paranal than found at La Palma. The reason for this discrepancy is due to the large wind shear within turbulent zones in the free atmosphere at Paranal, resulting in a dispersion of velocity vectors for the turbulent zone. The model does not have sufficient vertical resolution to resolve the velocity dispersion that is measured by the Stereo-SCIDAR.
   
\section{Conclusions}
\label{sect:conclusions}
In this publication we introduce the ESO Stereo-SCIDAR and show the statistical results which describe the Paranal site. We also compare the results to several other dedicated atmospheric characterisation instruments and models, namely the MASS-DIMM, Surface-Layer SLODAR and the ECMWF wind velocity forecast. This information is of interest to existing and future instruments for the Very Large Telescope at Cerro Paranal, Chile, as well as for the future Adaptive Optics instrumentations for the 38~m European Extremely Large Telescope which is being constructed approximately 20~km away.

We show high correlations between the Stereo-SCIDAR and the other instruments, with Pearson correlation coefficients between 0.60 (MASS-DIMM isoplanatic angle) and 0.84 (MASS-DIMM seeing). The median seeing, coherence time and isoplanatic angle is found to be 0.64\as, 4.18~ms and 1.75\as respectively. We also examine the fraction of the turbulence strength in the ground layer up to various altitudes. We find that the median ground layer fraction up to 300~m, 600~m, 900~m and 1200~m is 0.25, 0.40, 0.45 and 0.49 respectively. The ground layer fraction is critical to the performance of wide-field Adaptive Optics instrumentation.

\section*{Acknowledgements}

FP7/2013-2016: The research leading to these results has received funding from the European Community's Seventh Framework Programme (FP7/2013-2016) under grant agreement number 312430 (OPTICON).

Horizon 2020: This project has received funding from the European Union's Horizon 2020 research and innovation programme under grant agreement No 730890. This material reflects only the authors views and the Commission is not liable for any use that may be made of the information contained therein.

This work was supported by the Science and Technology Funding Council (UK) (ST/L00075X/1). DL acknowledges support from the UK Programme for the European Extremely Large Telescope (ST/N002660/1). OJDF acknowledges the support of STFC (ST/N50404X/1).

We also acknowledge ECMWF for access to the weather forecast data through the MARS access system.

This research made use of python including numpy and scipy \citep{numpy}, matplotlib \citep{matplotlib} and Astropy, a community-developed core Python package for Astronomy \citep{astropy}. We also made use of the python AO utility library `AOtools'.




\bibliographystyle{mnras}
\bibliography{Mendeley}


\bsp	
\label{lastpage}
\end{document}